\documentclass[12pt]{iopart}
\usepackage{iopams}
\usepackage{graphicx}

\begin{document}

\title{Is there life inside black holes?}

\author{V I Dokuchaev}

\address{Institute for Nuclear Research of the
Russian Academy of Sciences \\
60th October Anniversary Prospect 7a, 117312 Moscow, Russia}
 \ead{dokuchaev@inr.ac.ru}

\begin{abstract}
Bound inside rotating or charged black holes, there are stable periodic planetary orbits, which
neither come out nor terminate at the central singularity. Stable periodic orbits inside black
holes exist even for photons. These bound orbits may be defined as orbits of the third kind,
following the Chandrasekhar classification of particle orbits in the black hole gravitational
field. The existence domain for the third kind orbits is rather spacious, and thus there is
place for life inside supermassive black holes in the galactic nuclei. Interiors of the
supermassive black holes may be inhabited by civilizations, being invisible from the outside.
In principle, one can get information from the interiors of black holes by observing their
white hole counterparts.
\end{abstract}


\maketitle

\section{Introduction}
\label{intro}

A voyage inside the Black Hole (BH) event horizon may be not finished in the central
singularity after a finite proper time of the traveler, but once again outside the event
horizon. However, due to the complicated internal BH geometry
\cite{MisThWh,chandra,Wald84,FrolNov98,Frolov89,Frolov90,Frolov96}, it would be an emergence in
the \emph{other} universe rather than a return to the traveler's native one. Is it possible to
live long inside the BH avoiding both the downfall to the central singularity and escaping to
another universe? To clarify this possibility we suppose that BH interiors are described by the
Kerr-Newman metric with a maximally extended global geometry
\cite{MisThWh,chandra,Wald84,FrolNov98,Carter66,Carter66b,Carter68}. These BHs are named the
eternal ones. The corresponding Carter-Penrose conformal diagram of the eternal black is an
infinite stairway of asymptotically flat space-times, connected by the one-way Einstein-Rosen
bridges. The entries and outlets of these Einstein-Rosen bridges are, respectively, the event
horizons of the black and white holes. We demonstrate below that living inside the eternal BHs
is possible in principle, if these BHs are rotating or charged and massive enough for weakening
the tidal forces and radiation of gravitational waves to acceptable level.

A tidal acceleration experienced by the object (e.\,g., some creature) with a linear size $h$
at a distance $r$ from a black hole with mass $M$ is of the order of $g_t\sim GMh/r^3$ (see
e.\,g. Chapter 32.6 in \cite{MisThWh}). Let us accept that this tidal acceleration $g_t$ is in
a comfortable range for living species, if it does not exceed the corresponding gravitational
acceleration on the surface of the Earth, $g_{\rm E}\simeq10$~m/sec$^2$. From equality
$g_t=g_{\rm E}$ we find the corresponding tidal radius $r_t\sim (MGh/g_{\rm E})^{1/3}$. For a
comfortable traveling inside black hole this tidal radius must be less than the black hole
event horizon radius, $r_t<r_+\simeq GM/c^2$. From this inequality we estimate a corresponding
minimal black hole mass:
\begin{equation}
 \label{Mmin}
 M>M_{\rm min}\sim\frac{h^{1/2}c^3}{g_{\rm E}^{1/2}G}\simeq6.5\,10^3\sqrt{h}M_\odot.
\end{equation}
The minimal black hole mass $M_{\rm min}$ in (\ref{Mmin}) is much more less than masses of the
supermassive black hole candidates in the galactic nuclei, $M\sim10^6-10^{10}\,M_\odot$, for
the living species with a linear size $h\sim1-10^2$~cm.

After traversing the BH event horizon at radius $r=r_+$, a traveler will appear in the
$T$-region \cite{ZelNov}, where his radial coordinate $r$ would become a temporal one and
inevitably diminishing towards the central singularity. The irresistible infall in the
$T$-region will finish soon after traversing the inner Cauchy horizon at $r=r_-<r_+$, which is
nonzero for the rotating or charged BH. The internal space-time domain $0<r<r_-$ between the
central singularity and the inner BH horizon is the $R$-region, where stationary observers may
exist just as anywhere on the planet Earth. This internal BH domain, hidden by the two horizons
from the whole external universe, is indeed a suitable place for safe inhabitation. The only
thing needed is to put your vehicle or your planet to a stable periodic orbit inside BH. We
discuss some specific properties of stable periodic orbits of planets and photons inside the
rotating charged BH, described by the Kerr-Newman metric. It is supposed that generic
properties of the Kerr-Newman metric are survived inside black holes in spite of a threat from
the perturbative instabilities \cite{SimPen,Gursel79,Gursel79b,Novikov80,ChanHar82,Dotti08}.

\section{Stable periodic orbits inside BH}

The most generic description for the motion of particles in the gravitational field of the
black holes is a test particle approximation. This approximation is self-consistent, if
particles are small and light enough with respect to, respectively, the characteristic size and
mass of the black hole, so that the back reaction can be neglected. The term ``test planets''
is used throughout the paper as a synonym of the ``test particles''.

Subramanyan Chandrasekhar \cite{chandra} designated two general types of test particle orbits
in the BH gravitational field: orbits of the {\sl first kind}, which are completely confined
outside the BH event horizon, and orbits of the {\sl second kind}, which penetrate inside the
BH. Here we also propose to distinguish orbits of a {\sl third kind}, which are completely
bound inside the BH, not escaping outside, nor infalling into the central singularity. The
bound orbits of the third kind inside the inner BH horizon were found by Ji\v{r}\'\i~
Bi\v{c}\'ak, Zden\v{e}k Stuchl{\'\i}k and Vladim{\'\i}r Balek \cite{Bicak89a,Bicak89b} for
charged particles around the rotating charged BHs  (see also \cite{Kagramanova09,Olevares11})
and by Eva Hackmann, Claus L\"ammerzahl, Valeria Kagramanova and Jutta Kunz
\cite{Kagramanova10} for neutral particles around rotating BHs (see also \cite{Pugliese11}).
The bound orbits of the third kind are periodic and stable if gravitational and electromagnetic
radiation is  neglected.

Geodesics equations for neutral test particles and photons and equations of motion for charged
particles in the Kerr-Newman metric were derived by B. Carter \cite{Carter68}. According to
these equations the motion of test particle with a mass $\mu$ and electric charge $\epsilon$ in
the background gravitational field of a BH with a mass $M$, angular momentum $J=Ma$ and
electric charge $e$ is completely defined by three integrals of motion: the total particle
energy $E$,  the azimuthal component of the angular momentum $L$ and the Carter constant $Q$,
related with a total angular momentum of the particle. The Carter constant is zero, $Q=0$, if
trajectories are confined in the BH equatorial plane. In particular, the total angular momentum
of the particle is $\sqrt{Q+L^2}$ in the case of nonrotating BH.

An orbital trajectory of test planet is governed in the Boyer-Lindquist coordinates
$(t,r,\theta,\varphi)$ by equations of motion \cite{Carter68,bpt72}:
\begin{eqnarray}
 \Sigma\frac{dr}{d\lambda} &=& \pm \sqrt{V_r}, \label{rmot} \\
 \Sigma\frac{d\theta}{d\lambda} &=& \pm\sqrt{V_\theta}, \label{thetamot} \\
 \Sigma\frac{d\varphi}{d\lambda} &=& L\sin^{-2}\theta+a(\Delta^{-1}P-E),
 \label{phimot} \\
 \Sigma\frac{dt}{d\lambda} &=& a(L-aE\sin^{2}\theta)+(r^2+a^2)\Delta^{-1}P,
\end{eqnarray}
where $\lambda=\tau/\mu$, $\tau$ --- is a proper time of particle and
\begin{eqnarray}
 V_r &=& P^2-\Delta[\mu^2r^2+(L-aE)^2+Q], \label{Vr} \\
 V_\theta &=& Q-\cos^2\theta[a^2(\mu^2-E^2)+L^2\sin^{-2}\theta],
 \label{Vtheta} \\
 P &=& E(r^2+a^2)+\epsilon e r -a L, \\
 \Sigma &=& r^2+a^2\cos^2\theta, \\\
 \Delta &=& r^2-2r+a^2+e^2.  \label{Delta}
 \end{eqnarray}
We will use mainly the normalized dimensionless variables and parameters: $r\Rightarrow r/M$,
$a\Rightarrow a/M$, $e\Rightarrow e/M$,  $\epsilon\Rightarrow \epsilon/\mu$, $E\Rightarrow
E/\mu$, $L\Rightarrow L/(M\mu)$, $Q\Rightarrow Q/(M^2\mu^2)$. The radius of BH event horizon
$r=r_+$ and the radius of BH inner horizon $r=r_-$ are both the roots of equation $\Delta=0$: \
$r_{\pm}=1\pm\sqrt{1-a^2-e^2}$.

The effective potentials $V_r$ and $V_\theta$ in (\ref{Vr}) and (\ref{Vtheta}) define the
motion of particles in $r$- and $\theta$-directions \cite{bpt72}. In particular, for a circular
orbit at some radius $r$, equations (\ref{Vr}) and (\ref{Vtheta}) give conditions:
\begin{equation}
 \label{circularorb}
V_r(r)=0, \quad V_r'(r)\equiv\frac{dV_r}{dr}=0.
\end{equation}
The circular orbits would be stable if $V_r''<0$, i.~e. in the maximum of the effective
potential. (Note that the effective potential $V_r$ is defined with the opposite sign in
contrast to the usual definition of physical potential). In the case of a rotating BH (with
$a\neq0$), a particle on the orbit with $r=const$ may additionally be moving in the latitudinal
$\theta$-direction. These nonequatorial orbits are called {\em spherical orbits}
\cite{Wilkins72}. The purely circular orbits will correspond to the particular case of
spherical orbits with parameter $Q=0$, completely confined in the BH equatorial plane.

The generic orbits of the third kind are nonequatorial and periodic with respect to the
separate coordinates: $r$, $\theta$ and $\varphi$. Namely, (1) the $r$-periodicity means that
the orbital radial coordinate $r$ oscillates with a period $T_r$ between the minimal (perigee)
and maximal (apogee) values $r_p<r<r_a$ . The values of $r_p$ and $r_a$ are defined by zeroes
(the bounce points) of the radial potential, $V_r(r_{p,a})=0$. \ (2) The $\theta$-periodicity
means that the latitude coordinate $\theta$ oscillates with a period $T_\theta$ between the
minimal and maximal values, $\pi/2-\theta_{\bf max}<\theta<\pi/2+\theta_{\bf max}$, where
$\theta_{\bf max}$ is maximum angle of latitude elevation relative to the equatorial plane at
$\theta=\pi/2$.The value of $\theta_{\bf max}$ is defined by zero (the bounce point) of the
latitude potential $V_\theta(\theta_{\bf max})=0$. At last, (3) the $\varphi$-periodicity means
that the azimuth coordinate $\varphi$ oscillates with a period $T_\varphi$ between some
$\varphi_0$ and $\varphi_0+2\pi$. These three periods are incommensurable, i.\,e, all ratios
$T_r/T_\theta/T_\varphi$ are nor the rational numbers. For this reason the $3D$ space orbit of
the particle is not closed (but still periodic with respect to the separate coordinates).

\subsection{Circular orbits inside nonrotating charged BH}
\label{BHRN}

\begin{figure}[t]
\begin{center}
\includegraphics[width=0.85\textwidth]{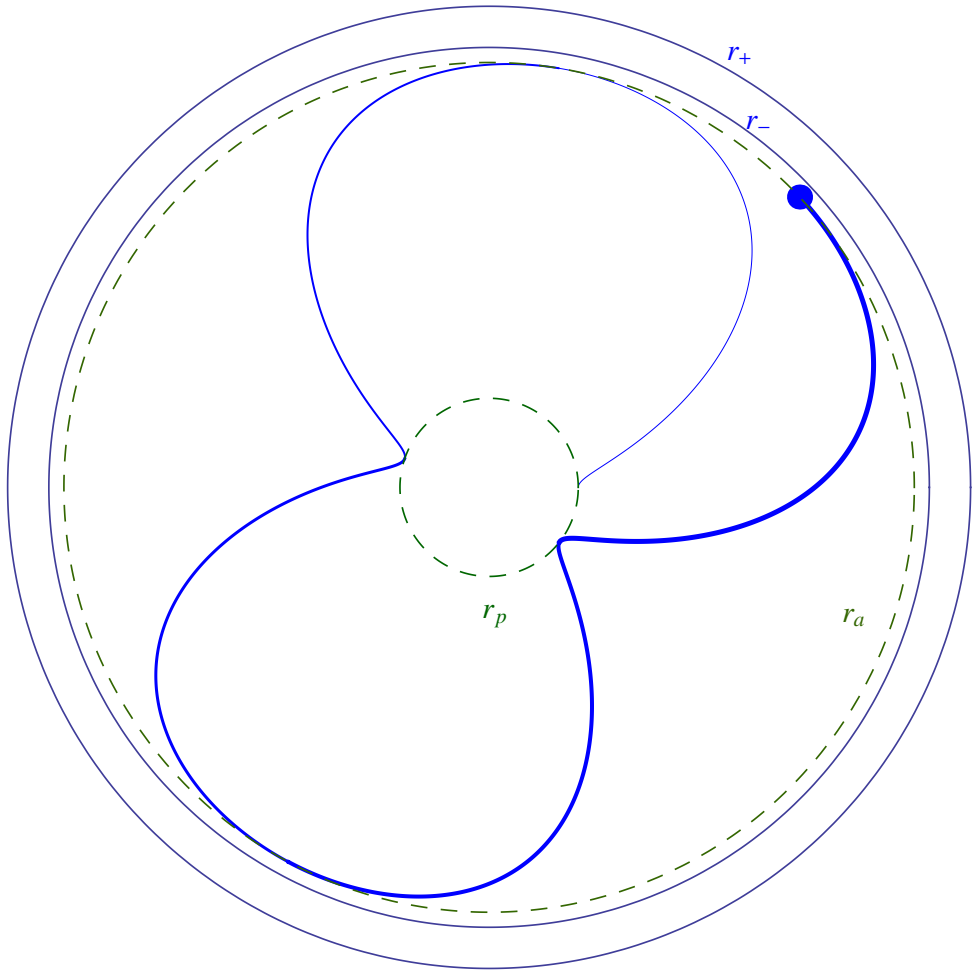}
\caption{The stable periodic orbit of planet with a mass $\mu$ and charge $\epsilon=-1.45\mu$
entirely inside the BH with a mass $M$ and charge $e=0.999M$. Orbit parameters: $E=1.5\mu$,
$L=0.2M\mu$, azimuthal and radial periods $(T_{\varphi},T_{r})=(14.9M,7.17M)$, the perigee and
apogee radii $(r_p,r_a)=(0.19M,0.92M)$.} \label{RNorbit}
\end{center}
\end{figure}

We first consider the Reissner-Nordstr\"om case of nonrotating charged BH. From the joint
resolution of equations (\ref{circularorb}) we find two pairs of solutions, respectively, for
the energy $E$ and angular momentum $L$ of massive particle with charge $\epsilon$ on the
circular orbit with radius $r$:
\begin{eqnarray}
E_{1,2}&=& \frac{\pm\Delta D_1-e\epsilon(r^2-4r+3e^2)}{2r(r^2-3r+2e^2)},
 \label{gl2circular} \\
L_{1,2}^2 &=& \frac{r^2}{r^2-3r+2e^2}\left[r-e^2+ \frac{e\epsilon\Delta(e\epsilon\pm
D_1)}{2(r^2-3r+2e^2)}\right].
 \label{l12circular}
\end{eqnarray}
where
\begin{equation}
 \label{D1}
 D_1^2=e^2(\epsilon^2+8)+4r(r-3).
\end{equation}
The stability condition $V_r''(r,E,L)<0$ for circular orbits inside the inner horizon,
$0<r<r_-$, is satisfied for the first pair of solution $(E_1,L_1,)$ in  equations
(\ref{gl2circular}) and (\ref{l12circular}), if $\epsilon>\mu$, and for the second pair
$(E_2,L_2,)$, if  $\epsilon<-\mu$. Figure~\ref{RNorbit} is an example of a stable periodic
orbit of a charged particle inside a nonrotating BH, calculated numerically from equations
(\ref{rmot}) and (\ref{phimot}), using a general formalism for motion in the central field
\cite{LL1}. The standard finite difference method was used in numerical integrations of
equations (\ref{rmot}), (\ref{thetamot}) and (\ref{phimot}) for the noncircular orbits with the
finite displacements of a proper time $\Delta\tau$ as the integration steps (or by using the
corresponding finite displacements of affine parameter $\lambda$ along the photon orbit).

\begin{figure}
\begin{center}
\includegraphics[width=0.8\textwidth]{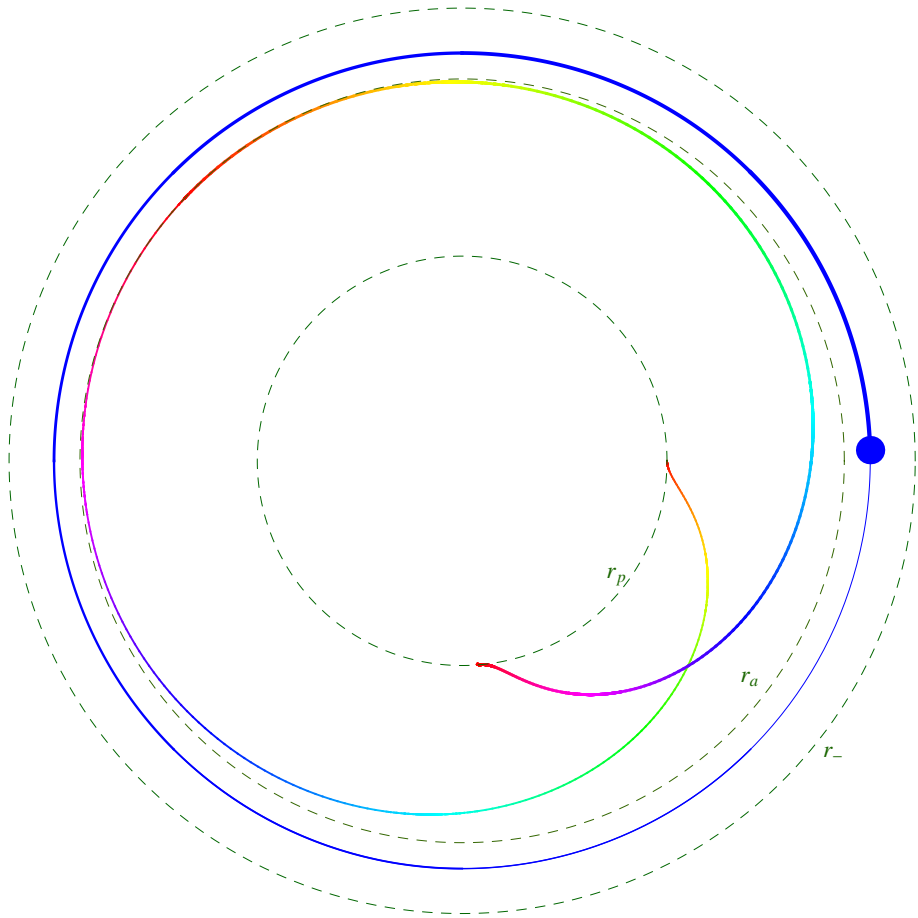}
\caption{The stable periodic equatorial photon orbit (colored curve viewed from the north pole)
with the impact parameter $b=1.53$ inside the BH with $a=0.75$ and $e=0.6$. Orbit parameters:
periods $(T_r,T_\varphi)=(2.7,2.1)$, perigee and apogee (dashed circles)
$(r_p,r_a)=(0.33,0.61)$. The external dashed circle is the inner horizon with $r=r_-=0.72$. The
blue circle is the circular planet orbit with the radius $r=0.65$, energy $E=10.5$ and impact
parameter $b=1.54$. The angular momentum of the black hole is directed to the north pole. The
thicknesses of orbital curves are growing with time. Note that the equatorial photon and
planet, starting at the azimuth angle $\varphi=0$, are both orbiting in opposite direction with
respect to BH rotation.
 \label{eqKNphoton}}
\end{center}
\end{figure}
\begin{figure}[t]
\begin{center}
\includegraphics[width=0.8\textwidth]{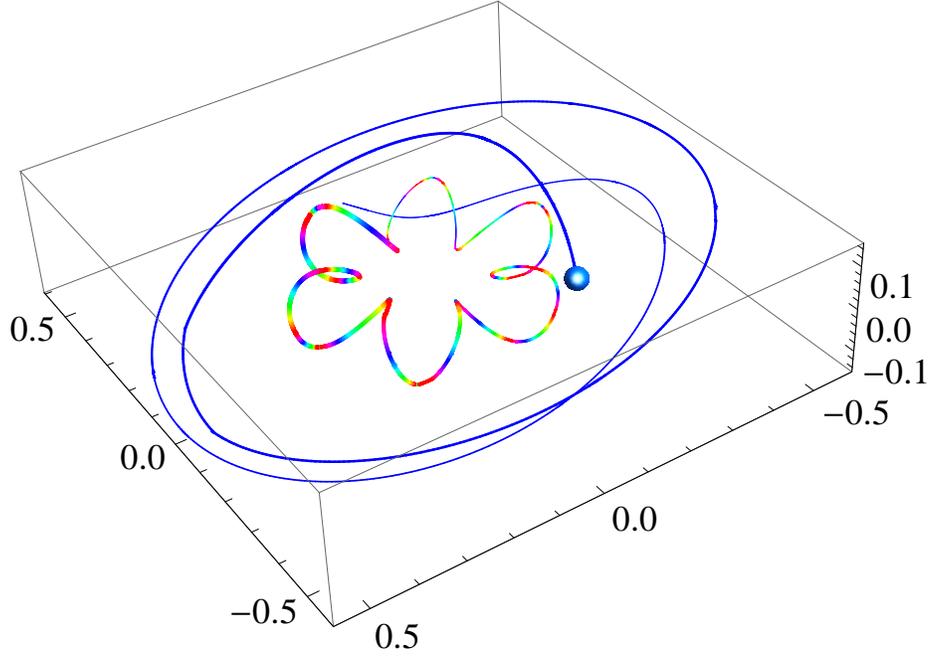}
\caption{{\em Outer curve}: the stable periodic orbit of planet with orbital parameters
$(E,L,Q)=(0.568,1.13,0.13)$, periods $(T_{\varphi},T_{r},T_\theta)=(1.63,3.70,1.17)$, apogee
and perigee radii $(r_p,r_a)=(0.32,0.59)$ and the maximum angle of latitude elevation relative
to the equatorial plane $\theta_{\rm max}=14.6^\circ$ inside the BH with $a=0.9982$ and
$e=0.05$. {\em Inner curve}: the stable periodic nonequatorial photon orbit with orbit
parameters: $(b,q)=(1.38,0.03)$, $(T_{\varphi},T_{r},T_\theta)=(2.95,0.49,0.33)$,
$(r_p,r_a)=(0.14,0.29)$ and $\theta_{\rm max}=10.1^\circ$ inside the same BH. The starting
parts of orbits are thin, while the ending parts are thick \label{FigFrame}}
\end{center}
\end{figure}
\begin{figure}
\begin{center}
\includegraphics[width=0.6\textwidth]{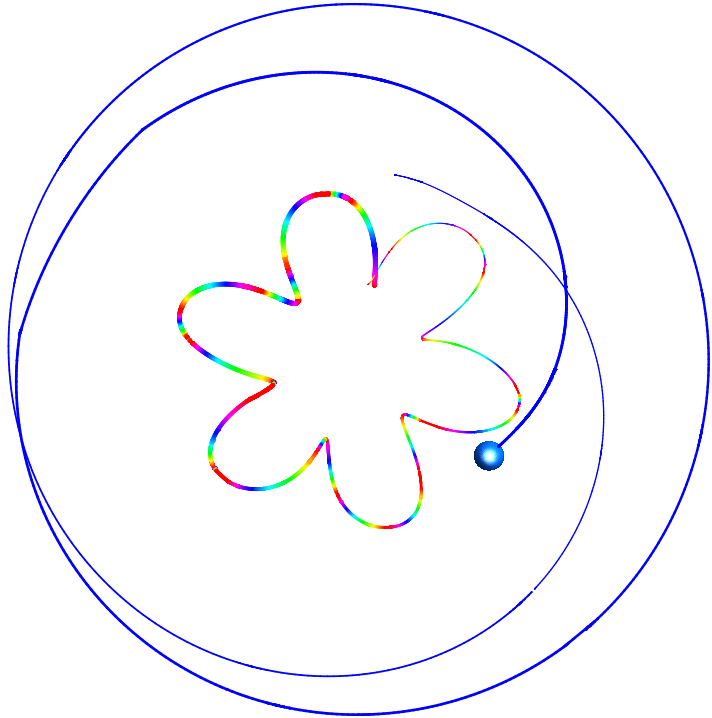}
\caption{The stable periodic orbits of photon and planet (shown in figure~\ref{FigFrame})
viewed from the coordinate frame north pole. The naked central singularity is glowing in the
center. The starting parts of orbits are thin and the finishing parts are thick.
\label{NoFrame}}
\end{center}
\end{figure}
\begin{figure}
\begin{center}
\includegraphics[width=0.9\textwidth]{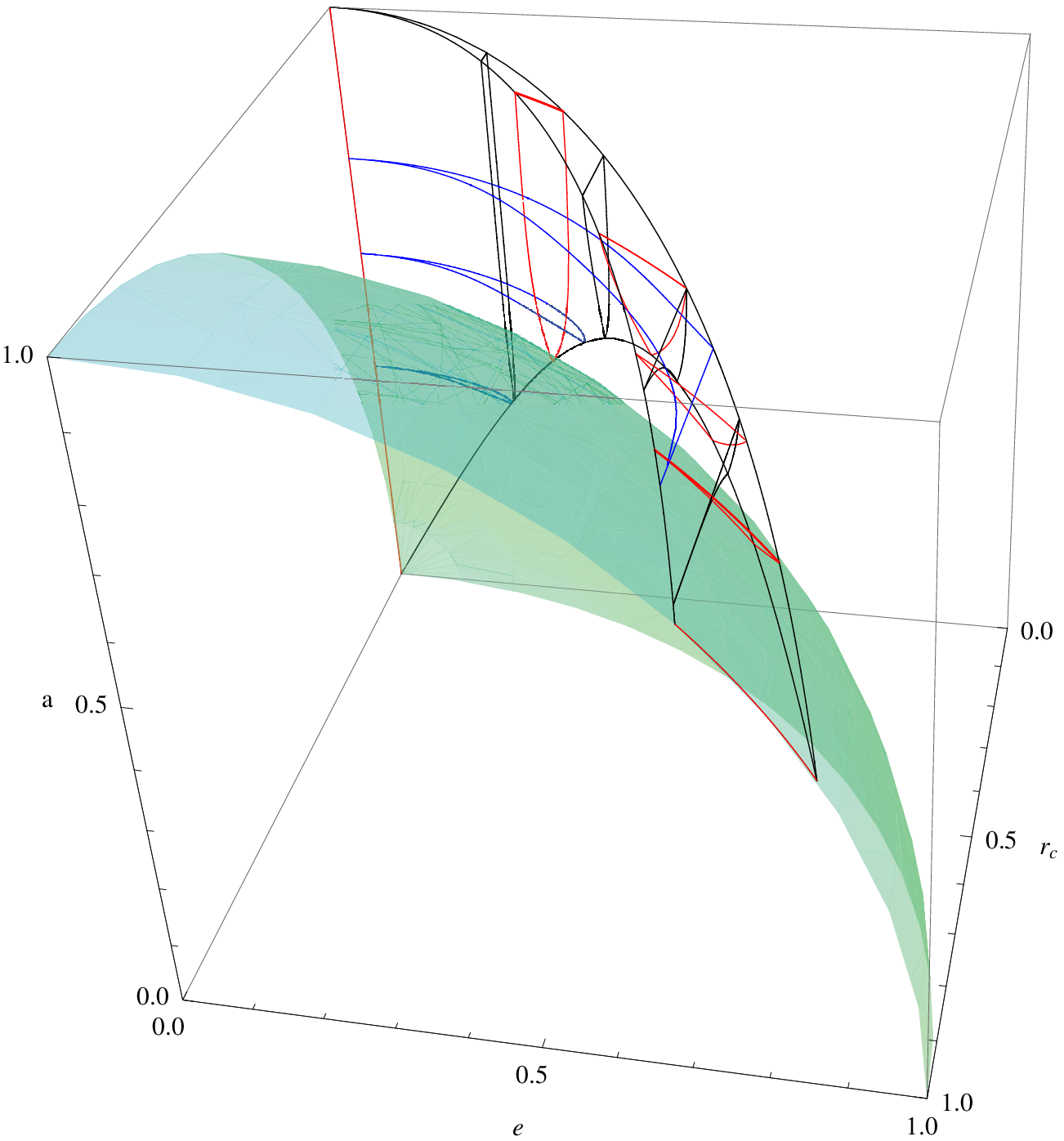}
\caption{$3D$ domain of existence (marked by the level curves $r=const$, $a=const$, $e=const$)
for the stable circular (equatorial) orbits of neutral particles inside the inner BH horizon (a
filled part of sphere). The rear filled surface corresponds to the ultrarelativistic limit or
circular photon orbits. The stable spherical (nonequatorial) orbits for massive particle exist
at radii, limited by the rear surface of circular photons and the inner horizon.
\label{domainex}}
\end{center}
\end{figure}

\subsection{Spherical orbits of neutral particles inside BH}
\label{BHKN}

In the Kerr-Newman case the stable circular orbits (and also the stable spherical ones) exist
inside the inner horizon not only for charged particles, but also for neutral ones
($\epsilon=0$) and photons ($\mu=0$). From equations (\ref{circularorb}) we find two pairs of
solutions for energy $E$ and azimuthal impact parameter  $b=L/E$ of neutral massive particle on
the spherical orbit:
\begin{eqnarray}
 \label{sphericalE}
 E^2_{1,2} &=& \frac{\mp2D_2+\beta_1r^2
 +a^2[2(r-e^2)\Delta-r^2(r-1)^2]Q}{r^4[(r^2-3r
 +2e^2)^2-4a^2(r-e^2)]}, \\
 b_{1,2} &=& \frac{\pm D_2 r - a^2(r-e^2)\{\beta_2 r
 + [a^2-r(r-e^2)]Q\}}
 {a(r-e^2)\{r[(\Delta-a^2)^2-a^2(r-e^2)]+a^2(1-r)Q\}},
 \nonumber
 \label{sphericalb}
\end{eqnarray}
where
\begin{equation}
\beta_1 = (r^2-3r+2e^2)(r^2-2r+e^2)^2 -a^2(r-e^2)[r(3r-5)+2e^2], \nonumber
\end{equation}
\begin{equation}
\beta_2 = e^4-a^2(r-e^2)+2e^2r(r-2)-r^2(3r-4),
 \label{beta2}
\end{equation}
\begin{equation}
 D_2^2 = [a(r-e^2)\Delta]^2[(r-e^2)r^4
 -r^2(r^2-3r+2e^2)Q+a^2Q^2]. \nonumber
\end{equation}
It can be shown that stable spherical orbits are realized for the first pair of solution
$(E_1,b_1)$ with $0<Q<Q_{\rm max}$, where $Q_{\rm max}$ is a root of the marginal stability
equation $V_r''=0$. All spherical orbits with $Q<0$ are unstable (see also \cite{Wilkins72}).
The analogous formulae for spherical orbits of charged particles in the Kerr-Newman case are
very cumbersome, and we are not reproducing them here.

\subsection{Spherical orbits of photons inside BH}
\label{SphericalPhotons}

The spherical photon orbit corresponds to the ultrarelativistic limit for massive particle
energy on the spherical orbit, $E\to\infty$. This limit is equivalent to the case $\mu=0$.
Photon orbit depends on two parameters, the azimuthal impact parameter $b=L/E$ and the
latitudinal (tangential) impact parameter $q=Q/E^2$. Equations (\ref{sphericalE}) in the
ultrarelativistic limit are very simplified:
\begin{equation}
 \label{b12}
 b_1 = \frac{a^2(1+r)+r(r^2-3r+2e^2)]}{a(1-r)},
 \qquad b_2=\frac{a^2+r^2}{a},
\end{equation}
\begin{equation}
 \label{q12}
 q_1 = \frac{r^2[4a^2(r-e^2)-(r^2-3r
 +2e^2)^2]}{a^2(1-r)^2},
 \qquad q_2=-\frac{r^4}{a^2}.
\end{equation}
Here the first pair of impact parameters, $(b_1,q_1)$, corresponds to stable spherical photon
orbits, while the second pair, $(b_2,q_2)$, to unstable ones. The stability condition
$V_r''\leq0$ for spherical photons with the first pair of impact parameters $(b_1,q_1)$ is
written in the form
\begin{equation}
a^2+e^2-r(r^2-3r+3)\leq0.
\end{equation}
From this inequality we find the upper limit of impact parameter $q_1>0$ for stable spherical
photons:
\begin{equation}
 q_1\leq\left(\frac{1-\delta^{1/3}}{a}\right)^2[3
 -4e^2-2(3-2e^2)\delta^{1/3}+3\delta^{2/3}],
\end{equation}
where $\delta=1-a^2-e^2$. The maximal allowable value for $q_1$ is reached for the extreme BH
with $a=\sqrt{1-e^2}\leq1/2$, $e\leq\sqrt3/2$:
\begin{equation}
q_{1,max}=4-a^{-2}\leq3.
\end{equation}
Figure~\ref{eqKNphoton} is an example of stable periodic equatorial photon orbit inside the
slightly charged but near extremely rotating BH. Figures~\ref{FigFrame} and \ref{NoFrame} are,
respectively, the examples of stable periodic planet and photon orbits inside a slightly
charged but near extremely rotating BH with the canonical specific angular momentum $a_{\rm
lim}=0.09982$ due to untwisting by a thin accretion disk \cite{Thorne74}.

\subsection{Circular orbits of photons inside BH}
\label{CircularPhotons}

The spherical orbits become circular ones in the particular case of $Q=0$. A corresponding
relation for circular photon orbits follows in the relativistic limit from equation
(\ref{sphericalE}) or from (\ref{b12}) and (\ref{q12}):
\begin{equation}
 \label{KNphotoncircular}
4a^2(r-e^2)=(r^2-3r+2e^2)^2.
\end{equation}
with two possible solutions for impact parameter
\begin{equation}
 \label{bphotoncircular}
 b_{1,2}=
 \frac{a\beta_2\pm r^2\sqrt{(r-e^2)\Delta^2}}{(r^2-2r+e^2)^2-a^2(r-e^2)},
\end{equation}
where $\Delta$ and $\beta_2$ are, respectively, from (\ref{Delta}) and (\ref{beta2}). In
(\ref{bphotoncircular}) the first solution $q_1$ (with a plus sign) corresponds to the stable
orbit, whereas $q_2$ corresponds to the unstable one. See in figure~\ref{domainex} the $3D$
domain of existence for the stable circular photons inside BH. These orbits exist at $e^2\leq
r\leq(4/3)e^2$, $a\neq0$, $0<e\leq\sqrt{3}/2$ and $0<b<5/2$. Note, that equatorial stable
orbits for planets and photons inside the Kerr BH ($e = 0$) are absent at all. The third kind
orbits inside the Kerr BH are only nonequatorial ones.

\section{Conclusions}
\label{Conclusion}

Inside the inner Cauchy horizon of the Kerr-Newman BH there are stable periodic orbits of
particles (planets) and photons (orbits of the third kind). In the case of a nonrotating
charged BH ($a = 0$) the stable periodic orbits exist only for particles with a large enough
charge. All the stable periodic planet and photon orbits inside the rotating and noncharged BH
($e = 0$) are nonequatorial. We hypothesize that civilizations of the third type (according to
Kardashev scale \cite{Kard64}) may live safely inside the supermassive BHs in the galactic
nuclei being invisible from the outside. Some additional highlighting during the night time
comes from eternally circulating photons. Yet, some difficulties (or advantages?) of a life
inside BH are worth mentioning, such as a possible causality violation
\cite{Carter66,Carter66b,Carter68} and the growing energy density and mass inflation in the
close vicinity of the Cauchy horizon \cite{SimPen,Gursel79,Gursel79b,Novikov80,ChanHar82}. The
existence of third kind orbits inside the event horizon may be verified or falsified in
principle (without the traveling into black holes) by the future observations of white holes.

\ack

Author acknowledges E.\,O. Babichev, V.\,A. Berezin, Yu.\,N. Eroshenko and I.\,D.~Novikov for
fruitful discussions. This work was supported in part by the Russian Foundation for Basic
Research grant 10-02-00635.

\end{document}